\documentclass[%
 amsmath,amssymb,
 aps,
]{revtex4-1}

\usepackage{graphicx}
\usepackage{dcolumn}
\usepackage{bm}

\begin{document}

\title{Topological entanglement entropy of the BTZ black hole}
\author{Jingbo Wang}
\email{ shuijing@mail.bnu.edu.cn}
\affiliation{Institute for Gravitation and Astrophysics, College of Physics and Electronic Engineering, Xinyang Normal University, Xinyang, 464000, P. R. China}
 \date{\today}
\begin{abstract}
In this paper, we calculate the topological entanglement entropy of the BTZ black hole and find that it coincides with that for fractional quantum Hall state. So the BTZ black holes have the same topological order with the fractional quantum Hall state. This fact indicate that the BTZ black holes have long-range entanglement.
\end{abstract}
\pacs{04.70.Dy,04.60.Pp}
 \keywords{topological entanglement entropy; topological order; generalized Hardy-Ramarujan formula;}
\bibliographystyle{unsrt}
\maketitle
\section{Introduction}
Thanks to the works of Event Horizon Telescope Collaboration \cite{eht1}, now we have the first picture of the black hole.\footnote{However in Ref.\cite{eht2} it was shown that we cannot exclude that the image might be something more exotic than a standard black hole.} It will accelerate the study about the black hole and the gravity theory. For black holes, there are still some difficult problems, a famous one is the information loss paradox \cite{info1,info2,info3}. An approach to solving this paradox is to find black hole-like objects in laboratory, such as acoustic black hole \cite{acoustic1} and so on  \cite{abh1}. One can study the Hawking radiation in those systems, and recently the thermal Hawking radiation was found in an analogue black hole \cite{abh2}.

In the previous works \cite{wangti1,wangti2}, the author claimed that the black hole can be considered as a kind of topological insulator. Roughly speaking, a topological insulator is a bulk insulator but has conducting boundary states. Based on this claim, we use the methods developed in topological insulator physics to study the problems in black hole physics \cite{wangbms2,wangplb1,wangbms4}. For Ba$\tilde{n}$ados-Teitelboim-Zanelli (BTZ) black holes in three dimensional spacetime this claim is tested in Ref.\cite{wangbms2,wangplb1}. The boundary modes on the horizon of the BTZ black hole can be described by two chiral massless scalar fields with opposite chirality \cite{whcft1}. This is the same as the topological insulator in three dimensional spacetime. The topological insulator has the $W_{1+\infty}$ symmetry group which contains the near horizon symmetry group of the BTZ black hole as a sub-group. For Kerr black holes in four dimension, the boundary modes can be described by boundary BF theory, which is also the same as four dimensional topological insulators. We give the microstates for BTZ black holes and Kerr black holes \cite{wangbms4}. Those microstates can account for the Bekenstein-Hawking entropy of those black holes.

An important question is: are those topological insulators integral or fractional? That is, can excited quasi-particles have fractional charges and statistics or not. This question is important since the fractional topological insulator has non-trivial topological order which results in many highly novel phenomena, including fractional statistics, protected gapless boundary excitations, and so on \cite{wen3}. In this paper, we calculate the topological entanglement entropy of the BTZ black hole and show that they are actually fractional topological insulators \cite{fti1,fti2} and have non-trivial topological order. It was claimed that the linearized gravity exhibits gapless topological order \cite{to1}, and from our result it is better to consider the black holes as gapped quantum liquid. The connection between the topological entanglement entropy of fractional quantum Hall state (FQHS) and black hole entropy was also discussed in Ref.\cite{he1}.

The paper is organized as follows. In section II, we give a brief review for the Laughlin states in FQHS. In section III, we analyse the BTZ black hole. Section IV is the conclusion. In the following we set $G=\hbar=c=1$.
\section{Review of the Laughlin states}
In three dimensional spacetime, the topological insulator can be considered as two copies of quantum Hall states. So firstly let us review the Laughlin states \cite{laughlin1} in fractional quantum Hall states \cite{tong1}. Laughlin states are the simplest FQHS with filling fractions $v=\frac{1}{m}$, where $m$ is an odd integer. The quantum Hall states contain no bulk gapless excitations but edge gapless excitations. Those low-energy edge excitations of quantum Hall states can be described by compact chiral boson field $\phi_L(t,\varphi)$ on a circle with coordinate $\varphi$ \cite{wen1}. The Lagrangian density is
\begin{equation}\label{1a}
  \mathcal{L}=\frac{1}{8\pi}[(\partial_t \phi_L)^2-(\partial_\varphi \phi_L)^2],
\end{equation}
with the chiral constraint
\begin{equation}\label{1b}
 (\partial_t-\partial_\varphi) \phi_L(t,\varphi)=0.
\end{equation}
On the circle, the chiral boson field can be expanded as
\begin{equation}\label{1}
  \phi_L(t,\varphi)=\phi_0+p_\phi(t+\varphi)+i\sum_{n\neq 0}\frac{1}{n}\alpha_n e^{-i n (t+\varphi)}.
\end{equation}
Those operators satisfy the Kac-Moody algebra
\begin{equation}\label{2a}
  [\alpha_n, \alpha_m]=n \delta_{n+m},\quad [\phi_0,p_\phi]=i,\quad others=0.
\end{equation}
Quantizing this boson field gives the following Hilbert space
\begin{equation}\label{2}
  \mathcal{H}=\mathcal{H}_{KM}\otimes \mathcal{H}_p,
\end{equation}
where $\mathcal{H}_{KM}$ is generated by the oscillator part $\alpha_n$, and represent the zero charge states, that is the phonon. The $\mathcal{H}_p$ is generated by the zero mode part $(\phi_0,p_\phi)$, and represent the charged states, that is the quasi-particles and quasi-holes.

For Laughlin states $v=\frac{1}{m}$, the zero mode part satisfies the quantized condition \cite{wen1}
\begin{equation}\label{2b}
p_\phi= \sqrt{m} n,\quad n \in N,
\end{equation}
so the boson field has the period
\begin{equation}\label{3}
  \phi_L(t,\varphi+2\pi)=\phi_L(t,\varphi)+2\pi \sqrt{m} n.
\end{equation}
It is more convenient to rescale the boson field to $\phi'_L=\sqrt{m} \phi_L$ which has period $2\pi n$. The Lagrangian density for this field is
\begin{equation}\label{3a}
  \mathcal{L}=\frac{m}{8\pi}[(\partial_t \phi'_L)^2-(\partial_\varphi \phi'_L)^2].
\end{equation}
This boundary field theory describe the boundary modes for the FQHS.

For fractional quantum Hall state with filling factor $v=\frac{1}{m}$, the entanglement entropy is given by
\begin{equation}\label{4}
  S=\alpha \frac{l}{\epsilon}-\gamma+O(\frac{1}{l}),
\end{equation}
where $\epsilon$ is the UV-cutoff, $l$ the length of the boundary and $\alpha$ a non-universal constant. The second term $\gamma$ is a universal term which is called `topological entanglement entropy' (TEE) \cite{tee1,tee2} and is a characteristic of the topological order of the state. Such a term indicates the existence of certain long-rang entanglement structure that originates from the topological nature of the system \cite{wen3}. For Laughlin state $v=\frac{1}{m}$ the TEE takes the form $\gamma=\frac{1}{2}\ln m$.
\section{The topological entanglement entropy of the BTZ black hole}
Now let us consider the BTZ black hole case. Similar to quantum Hall states, the black hole has also boundary excitations. Those boundary degrees of freedom can be described by two chiral massless scalar fields $\Psi_L$ (for left-moving) and $\Psi_R$ (for right-moving) \cite{whcft1}. From this point of view, the BTZ black hole can be considered as two copies of quantum Hall states, or a quantum spin Hall state. The metric of the BTZ black hole is \cite{btz1}
\begin{equation}\label{5}
    ds^2=-N^2 dv^2+2 dv dr+r^2 (d\varphi+N^\varphi dv)^2,
\end{equation}
where $N^2=-8 M+\frac{r^2}{l_0^2}+\frac{16 J^2}{r^2}, N^\varphi=-\frac{4 J}{r^2}$, and $M,J$ are the mass and angular momentum of the BTZ black hole respectively, $l_0$ the radius of the AdS spacetime. The black hole has the event horizon at $r=r_+$ and the inner horizon at $r=r_-$ with
\begin{equation}\label{6}
  r^2_\pm=4 M l_0^2 (1\pm \sqrt{1-\frac{J^2}{M^2 l_0^2}}).
\end{equation}

The degrees of freedom on the horizon can be described by two compact chiral scalar fields on the horizon. The action for right-moving section $\Psi_R$ is given by \cite{whcft1}
\begin{equation}\label{10}\begin{split}
    I_R=\frac{k l_0}{4\pi}\int_{\Delta} dv d\varphi[(\partial_v \Psi_R)^2- \frac{1}{l_0^2}(\partial_{\varphi} \Psi_R)^2],\quad k=\frac{l_0}{4},
\end{split}\end{equation}
with constraints \cite{wangplb1}
\begin{equation}\label{11}\begin{split}
 (\partial_v+\frac{1}{l_0}\partial_\varphi) \Psi_R(v,\varphi)=0,\\
 \Psi_R(v,\varphi+2\pi)=\Psi_R(v,\varphi)-\frac{\pi(r_++r_-)}{l_0}.
\end{split}\end{equation}
We also rescale the field $\Psi'_R=\frac{2 l_0}{(r_++r_-)} \Psi_R$ to have period $2\pi$. The action for the field $\Psi'_R$ is
\begin{equation}\label{12}\begin{split}
    I_R=\frac{k l_0}{4\pi}(\frac{r_++r_-}{2 l_0})^2\int_{\Delta} dv d\varphi[(\partial_v \Psi'_R)^2- \frac{1}{l_0^2}(\partial_{\varphi} \Psi'_R)^2].
\end{split}\end{equation}
The effective metric for this action is $ds^2=-dv^2+l_0^2 d\varphi^2$ which is a circle with radius $l_0$. Later we will show that it is better to define the theory on a circle with radius $r_R=\frac{l_0^2}{r_++r_-}$, which can be arrived by a conformal transformation $ds_R^2=\Omega_R^2 ds^2$ with $\Omega_R=\frac{l_0}{r_++r_-}$. The metric is
\begin{equation}\label{12a}
  ds_R^2=\Omega_R^2 (-dv^2+l_0^2 d\varphi^2)=-dv_R^2+r_R^2 d\varphi^2,
\end{equation}
with $v_R= \Omega_R v, r_R= \Omega_R l_0.$

The action becomes
\begin{equation}\label{13}\begin{split}
    I_R=\frac{k}{16\pi}(\frac{r_++r_-}{l_0})^2\int_{\Delta} dv_R  d\varphi[(\partial_{v_R} \Psi'_R)^2- \frac{1}{r_R^2}(\partial_{\varphi} \Psi'_R)^2]\\
    =\frac{k \Omega_R}{16\pi}(\frac{r_++r_-}{l_0})^2 \int_{\Delta} dv  d\varphi[(\partial_{v_R} \Psi'_R)^2- \frac{1}{r_R^2}(\partial_{\varphi} \Psi'_R)^2]\\
    =\frac{k_R}{4\pi}\int_{\Delta} dv  d\varphi[(\partial_{v_R} \Psi'_R)^2- \frac{1}{r_R^2}(\partial_{\varphi} \Psi'_R)^2]
\end{split}\end{equation}
with $k_R=\frac{k \Omega_R}{4}(\frac{r_++r_-}{l_0})^2=\frac{r_+ + r_-}{16}$. This action coincide with (\ref{3a}) that describe the fractional quantum Hall state with filling factor $v=\frac{1}{m}=\frac{1}{2k_R}.$

The compact boson field has mode expansion
\begin{equation}\label{14}
  \Psi'_R(v_R,\varphi)= \Psi_{R0}+p_R (v_R-r_R \varphi)+\sqrt{\frac{1}{k_R r_R}}\sum_{l>0}\sqrt{\frac{1}{2 \omega_{Rl}}}[a_l e^{-i(\omega_{Rl} v_R-k_l \varphi)}+a^+_l e^{i(\omega_{Rl} v_R-k_l \varphi)}],
\end{equation}
where $\omega_{Rl}=\frac{l}{r_R},k_l=l$ and $A=2\pi r_R$ is the length of the circle. Since the compact boson field $\Psi'_R$ has period $2\pi$, one can get
\begin{equation}\label{21a}
 p_R=\frac{1}{r_R}=\frac{r_++r_-}{l_0^2}.
\end{equation}

The Hamiltonian for this scalar field can be given by
\begin{equation}\label{22}\begin{split}
  H_R=\frac{k_R}{4\pi}\oint d\varphi\sqrt{-g} ((\partial_0  \Psi'_R)^2+ (\frac{\partial_1  \Psi'_R}{r_R})^2)\\
  =k_R r_R p_R^2+\sum_{l> 0} \frac{l}{r_R}\hat{a}_l^+ \hat{a}_l=\frac{(r_++r_-)^2}{16 l_0^2}+\sum_{l> 0} \frac{l}{r_R}\hat{n}_l,
 \end{split}\end{equation}
where $\hat{n}_l=\hat{a}_l^+ \hat{a}_l$ is the number operator.

The re-scaled chiral compact boson field $\Psi'_L=\frac{2 l_0}{(r_+-r_-)}\Psi_L$ on left-moving sector has the similar structure. It moves on the cylinder with effective metric
\begin{equation}\label{24}
  ds_L^2=\Omega_L^2 (-dv^2+l_0^2 d\varphi^2)=-dv_L^2+r_L^2 d\varphi^2,
\end{equation}
where the radius is $r_L=\frac{l_0^2}{r_+-r_-}$ and $\Omega_L=\frac{l_0}{r_+-r_-}.$

The boson field has expansion
\begin{equation}\label{25}
  \Psi'_L(v_L,\varphi)= \Psi_{L0}+p_L (v_L+r_L \varphi)+\sqrt{\frac{1}{k_L r_L}}\sum_{l<0}\sqrt{\frac{1}{2 \omega_{Ll}}}[a_l e^{-i(\omega_{Ll} v_L-k_l \varphi)}+a^+_l e^{i(\omega_{Ll} v_L-k_l \varphi)}],
\end{equation}
where $\omega_{Ll}=\frac{l}{r_L}.$
The compact boson field $ \Psi'_L$ also has period $2\pi$, which gives
\begin{equation}\label{25a}
 p_L=\frac{1}{r_L}=\frac{r_+-r_-}{l_0^2}.
\end{equation}

The Hamiltonian for this scalar field can be given by
\begin{equation}\label{26}\begin{split}
  H_L=\frac{k_L}{4\pi}\oint d\varphi\sqrt{-g} ((\partial_0  \Psi'_L)^2+ (\frac{\partial_1  \Psi'_L}{r_L})^2)\\
  =k_L r_L p_L^2+\sum_{l<0} \frac{-l}{r_L}\hat{a}_l^+ \hat{a}_l=\frac{(r_+-r_-)^2}{16 l_0^2}+\sum_{l<0} \frac{-l}{r_L}\hat{n}_l.
 \end{split}\end{equation}

The scalar field $ \Psi'_{R/L}(v,\varphi)$ can be considered as collectives of harmonic oscillators, and a general quantum state can be represented as $|p_R,;\{n_{l>0}\}>$ and $|p_L,;\{n_{l<0}\}>$ where $p_R,p_L$ are zero mode parts, and $\{n_l\}$ the oscillating parts. The parameters of the BTZ black hole can be obtained from the zero-mode of those two chiral boson fields with
\begin{equation}\label{27}\begin{split}
  H=H_{R0}+H_{L0}=\frac{(r_++r_-)^2}{16 l_0^2}+\frac{(r_+-r_-)^2}{16 l_0^2}=\frac{r^2_++r^2_-}{8 l_0^2}=M,\\
  J/l_0=H_{R0}-H_{L0}=\frac{(r_++r_-)^2}{16 l_0^2}-\frac{(r_+-r_-)^2}{16 l_0^2}=\frac{r_+r_-}{4 l_0^2}.
\end{split}\end{equation}

The calculation of the entropy for the BTZ black hole is as follows. For the BTZ black hole with parameters $(M,J)$, the oscillating parts satisfy the constraints
\begin{equation}\label{29}
 \sum_{l>0} \frac{l}{r_R} n^R_l+\sum_{l<0} \frac{-l}{r_L} n^L_l=M,\quad  \sum_{l>0} \frac{l}{r_R} n^R_l-\sum_{l<0} \frac{-l}{r_L} n^L_l=J/l_0.
\end{equation}

They are equivalent to
\begin{equation}\label{30}
 \sum_{l>0} l n^R_l=\frac{1}{2}(M+J/l_0) r_R=k_R,\quad  \sum_{l<0} -l n^L_l=\frac{1}{2}(M-J/l_0) r_L=k_L.
\end{equation}
The partitions of a positive integer number $N$ is given by the famous Hardy-Ramarujan formula
\begin{equation}\label{41}
  p(N)\simeq \frac{1}{4\sqrt{3}N}\exp(2\pi \sqrt{\frac{N}{6}}).
\end{equation}
Now we assume that every $n^R_k(n^L_k)$ carry $c_R=24 k_R=\frac{3(r_+ + r_-)}{2}(c_L=24 k_L=\frac{3(r_+ - r_-)}{2})$ labels, that is, every harmonic oscillator has $c_R(c_L)$ possible polarizations.
We denote $p_c(N)$ the number of partitions of $N$ into integers that can carry $c$ labels. Then a general Hardy-Ramarujan formula exist \cite{hw1,string1}
\begin{equation}\label{42}
  p_c(N)\simeq \frac{1}{\sqrt{2}}(\frac{c}{24})^{(c+1)/4}N^{-(c+3)/4}\exp(2\pi \sqrt{\frac{c N}{6}}).
\end{equation}
The entropy is related to the logarithm of this function, thus
\begin{equation}\label{43}
  \ln p_c(N)\simeq 2\pi \sqrt{\frac{c N}{6}}-\frac{c+1}{4}\ln \frac{24 N}{c}-\frac{1}{2}\ln (2N).
\end{equation}
For $c=24 N$, which is in our case, one can get
\begin{equation}\label{44}
  \ln p_c(N)\simeq 4\pi N-\frac{1}{2}\ln (2N).
\end{equation}

For the constraints (\ref{30}) one can get
\begin{equation}\label{31}\begin{split}
  S_R=4\pi k_R-\frac{1}{2}\ln (2k_R)=\frac{\pi}{4}(r_++r_-)-\frac{1}{2}\ln \frac{r_++r_-}{8},\\
  S_L=4\pi k_L-\frac{1}{2}\ln (2k_L)=\frac{\pi}{4}(r_+-r_-)-\frac{1}{2}\ln \frac{r_+-r_-}{8}.
\end{split}\end{equation}
and the total entropy is
\begin{equation}\label{32}
  S=S_R+S_L=\frac{2\pi r_+}{4}-\frac{1}{2}\ln \frac{r^2_+-r^2_-}{64}.
\end{equation}
The logarithmic term in (\ref{31}) for right sector is the same with the topological entanglement entropy for fractional quantum Hall state with filling factor $v=\frac{1}{2k_R}$, that is,
\begin{equation}\label{33}
  \gamma=\ln \sqrt{m}=\frac{1}{2}\ln 2 k_R.
\end{equation}
The same is true for left sector. So we can say that the right (left) sector of the BTZ black hole has the same topological order with the fractional quantum Hall state with filling factor $v=\frac{1}{2k_R}$ ($v=\frac{1}{2k_L}$).

\section{Discussion and Conclusion}
In this paper, we claim that the BTZ black hole has the same topological order as fractional quantum Hall state. They are in the same universal class. Topological order are nothing but the pattern of many-body long-range entanglement \cite{wen3}. Thus the black hole has long-range entanglement. Entanglement is very useful to build the spacetime geometry \cite{van1}. This gives an approach to study the black hole and gravity with the methods developed in the condensed matter physics. It is also interesting to investigate if black holes can support non-abelian anyons which can be used to build the topological quantum computer \cite{tqc1}.

In AdS spacetime, there exist the Hawking-Page phase transition between AdS-Schwarzschild black holes at high temperature and thermal AdS gas at low temperature \cite{hp1}. This phase transition is first order, which can be considered as liquid/solid phase transition. This transition is also possible for non-rotating BTZ black holes \cite{hp2,hp3}. We can analyse this phase transition from our approach. For non-rotating BTZ black holes, the central charge is $c_R=c_L=\frac{3 r_+}{2}$. On the other hand, it is well known that the central charge at infinity is $c=\frac{3 l_0}{2 }$ \cite{bh1}. Due to the red-shift, we consider the near horizon region as ultra-violet (UV) region and infinity as the infra-red (IR) region. Due to the $c-$theorem \cite{cth1}, one should have
\begin{equation}\label{34}
  c_R\geq c\Rightarrow r_+\geq l_0,
\end{equation}
which is just the condition for the stable of the non-rotating BTZ black hole. One can also understand this phase transition from the FQHE. Actually the FQH states can undergo a first order phase transition to Wigner crystal when the density of the electron is low \cite{wc1}. So the BTZ black hole-thermal gas phase transition is similar to FQHE-Wigner crystal transition.
\acknowledgments
 This work is supported by Nanhu Scholars Program for Young Scholars of XYNU.

\bibliography{cc1}

\end{document}